\documentclass [12pt,elsart,epsfig]{article}     
\usepackage{epsfig}
\begin{document}

\begin {center}
{\Large \bf Partial  Wave Analysis of $\bar pp \to \pi ^- \pi ^+$, $\pi ^0 \pi ^0$,
$\eta \eta$ and $\eta \eta '$}
\vskip 5mm

{A.V. Anisovich$^c$, C.A. Baker$^a$, C.J. Batty$^a$, D.V. Bugg$^b$, A. Hasan$^b$, 
C. Hodd$^b$,
J. Kisiel$^d$, V.A. Nikonov$^c$, A.V. Sarantsev$^c$, V.V. Sarantsev$^c$, 
B.S.~Zou$^{b}$ \footnote{Now at IHEP, Beijinj 100039, China} \\
{\normalsize $^a$ \it Rutherford Appleton Laboratory, Chilton, Didcot OX11 0QX,UK}\\
{\normalsize $^b$ \it Queen Mary and Westfield College, London E1\,4NS, UK}\\
{\normalsize $^c$ \it PNPI, Gatchina, St. Petersburg district, 188350, Russia}\\
{\normalsize $^d$ \it University of Silesia, Katowice, Poland}\\ 
[3mm]}
\end {center}

\begin{abstract}
A partial wave analysis is presented of Crystal Barrel data on
$\bar pp \to \pi ^0 \pi ^0$,
$\eta \eta$ and $\eta \eta '$ from 600 to 1940 MeV/c, combined with earlier
data on $d\sigma /d\Omega$ and $P$ for $\bar pp \to \pi ^-\pi^+$.
The following $s$-channel $I=0$ resonances are identified:
(i) $J^{PC} = 5^{--}$ with mass and width ($M$, $\Gamma$) at ($2295 \pm 30$,
$235 ^{+65}_{-40}$) MeV,
(ii) $J^{PC} = 4^{++}$ at ($2020\pm 12$,  $170 \pm 15$) MeV and
($2300 \pm 25$, $270 \pm 50$) MeV,
(iii) $^3D_3$ $J^{PC} = 3^{--}$ at ($1960 \pm 15$, $150\pm 25$) MeV and
($2210 \pm 40$, $360 \pm 55$) MeV,
and a $^3G_3$ state at ($2300 ^{+50}_{-80}, 340 \pm 150)$ MeV,
(iv) $J^{PC} = 2^{++}$ at ($1910 \pm 30$, $260\pm 40$) MeV,
($2020 \pm 30$, $275 \pm 35$) MeV,
($2230 \pm 30$, $245 \pm 45$) MeV, and
($2300 \pm 35$, $290 \pm 50$) MeV,
(v) $J^{PC} = 1^{--}$ at ($2005 \pm 40$, $275\pm 75$) MeV, and
($2165 \pm 40$, $160 ^{+140}_{-70}$) MeV, and
(vi) $J^{PC} = 0^{++}$ at ($2005 \pm 30$, $305 \pm 50$) MeV,
($2105 \pm 15$, $200\pm 25$) MeV, and
($2320 \pm 30$, $175 \pm 45$) MeV.
In addition, there is a  less well defined $6^{++}$ resonance at $2485 \pm 40$
MeV, with $\Gamma = 410 \pm 90$ MeV.
For every $J^P$, almost all these resonances lie on well defined linear
trajectories of mass squared v. excitation number.  The slope is
$1.10 \pm 0.03 $ Gev$ ^2$ per excitation.

The $f_0(2105)$ has strong coupling to
$\eta \eta$, but much weaker coupling to $\pi ^0 \pi ^0$.
Its flavour mixing angle between $q \bar q$ and $s \bar s$ is
$(59-71.6)^{\circ}$, i.e. dominant decays to $s\bar s$.
Such decays and its strong production in $\bar pp$ interactions
strongly suggest exotic character.

\end{abstract}


\section {Introduction}
In an accompanying paper [1], data are presented from the Crystal Barrel
experiment on $\bar pp \to \pi ^0 \pi ^0$, $\eta \eta$ and $\eta \eta '$
at nine beam momenta from 600 to 1940 MeV/c; this corresponds to the mass range
1960--2410 MeV.
The objective of the present paper is to add these new results to
earlier data on
$\bar pp \to \pi ^- \pi ^+$ and do a combined partial wave analysis.
Data on $d\sigma /d\Omega$ were reported by Eisenhandler et al. [2] and on
polarisation $P$ by Carter et al. [3] from $990$ to 2430 MeV/c (masses up to
2580 MeV).
Further data on both $d\sigma /d\Omega$ and $P$ from 360 to 1550 MeV/c were
reported by Hasan et al. [4].
The polarisation data play a vital role in the analysis, since they separate
cleanly states with $L = J \pm 1$, e.g. $^3P_2$ and $^3F_2$.
These data help greatly in consolidating the analyses of $\pi ^0 \pi ^0$,
$\eta \eta$ and $\eta \eta '$ data and help identify resonances.
The data of Hasan et al. are particularly valuable, since they cover the
important mass range down to 1910 MeV.

There have been several earlier amplitude analyses of the data on $\bar pp \to
\pi ^- \pi ^+$ [5-11].
We find quite good agreement with these papers, though we locate
additional features, noteably four $2^+$ states.
The earlier analyses were restricted to the $\pi \pi$ channel only.
One of our primary
objectives is to include $\eta \eta$ and $\eta \eta '$ data, so as to examine
the SU(3) character of the fitted resonances.

We also use data on $4^+$ states observed in $\bar pp \to \eta
\pi ^0 \pi ^0$ [12].
They provide further constraints on the relative sign and magnitudes of
$^3H_4$ and $^3F_4$ amplitudes.

The analysis reported here is in terms of $s$-channel resonances up to $J^{PC}
= 6^{++}$, plus backgrounds which are needed only in the low partial waves $0^{++}$ and $1^{--}$.
There are a number of reasons for this approach in terms of resonances.
Firstly, we find that the well
known $f_4(2050)$  makes a large and unavoidable contribution.
It fits to a mass and width slightly lower than averages quoted by the
Particle Data Group (PDG) [13].
Because we need to make use of its fitted mass, we shall hereafter
refer to it as $f_4(2020)$.
Once this resonance is introduced, it defines phases over a considerable mass
range. Relative phases with other partial waves are accurately determined
and it becomes impossible to avoid introducing
further resonances into all other partial waves. This is not surprising, since
resonances are predicted in this mass range by extrapolation from experience at
lower masses. A second related point concerns the
polarisation.
It lies close to +1 over much
of the angular range at many momenta and  inevitably requires large
imaginary parts in many partial wave amplitudes.
The quantity $Pd\sigma /d\Omega$ is proportional to the imaginary part of
the interference between partial waves.
Differential cross sections
measure the real parts of the same interferences.
In order to achieve consistency between the principle of analyticity
and these real and imaginary parts of interferences, we find resonance
behaviour in all partial waves unavoidable.
This is what was found in all earlier analyses except that of Kloet and
Myrher [11], which focussed on the narrow mass range below a momentum of
1 GeV/c.

\section {Procedures and Formulae}
For completeness, we repeat the standard formulae used in earlier work.
The differential cross section may be expressed in terms of spin-flip
($F_{+-}$) and non-flip ($F_{++}$) helicity amplitudes:
\begin {eqnarray}
d\sigma /d\Omega &=& |F_{++}|^2 + |F_{+-}|^2 \\
F_{++} &=& \frac {1}{4}\sum_{J=0}^{J_{max}} (2J + 1)f^J_{++}P_J(\cos \theta ) \\
F_{+-} &=& \frac {1}{4}\sum_{J=1}^{J_{max}} \frac {(2J + 1)}{\sqrt {J(J + 1)}}
f^J_{+-}P^1_J(\cos \theta ).
\end {eqnarray}
Here $P_J^m$ are Legendre polynomials.
Partial helicity amplitude $f^J$ are defined in terms of angular momentum
partial waves $T_{L,J}$ according to:
\begin {eqnarray}
\sqrt {2J + 1}f^J_{++} = \sqrt {J}T_{J-1,J} - \sqrt {J + 1}T_{J+1,J} \\
\sqrt {2J + 1}f^J_{+-} = \sqrt {J + 1}T_{J-1,J} + \sqrt {J}T_{J+1,J}.
\end {eqnarray}

Our procedure is to  express the $T_{L,J}$ as sums over resonances, up to
5 for each $J$ value:
\begin {equation}
T_{L,J} = \sum _{i=1}^5 \frac {G _i B_L(p)B_J(q)\exp (i\phi _i)}
{M_i^2 - s - iM_i\Gamma _i}.
\end {equation}
Here $B_L$ are standard Blatt-Weisskopf centrifugal barrier factors for angular
momentum $L$ in terms of momentum $p$ in the entrance channel $\bar pp$ and
$q$ in the exit meson channel.
Explicit formulae are given by Chung [14].
They guarantee the correct threshold behaviour in the $\bar pp$ channel.
The $G_i$ are real coupling constants and $\phi _i$ are phases for each
resonance, arising from final state interactions.

We now comment in general terms on this approach.
Firstly, resonances overlap and interact to some degree, e.g. via common decay
channels. In principle, a K-matrix approach is desirable, in order to
investigate such interactions. In reality, this is impracticable at present,
because of the large number of open channels, and almost complete lack of
information about many of them. A practical point is that
the speed of convergence of the fitting procedure is helped greatly by using
as basis states the eigenstates of the scattering amplitudes;
these are T-matrix poles.
We therefore view the resonances fitted to the data as a parametrisation
from which further analysis can unfold the physics
content. A first step is to find how many resonances are required
and what their properties are.

A second concern is that $t$-channel exchanges also undoubtedly contribute.
What effect are they likely to have, if any?
In each partial wave, they lead to left-hand singularities which are very
distant;
the left-hand cut opens below $s \simeq 1$ GeV$^2$.
For $J \le 3$, there are further resonances lying between the mass range
discussed here and the left-hand cuts.
Any $t$-channel amplitude
will acquire the phase dictated by $s$-channel resonances in each partial
wave, via rescattering within the present mass range.
The effect of any $t$-channel exchange is therefore to introduce some
slow $s$-dependence into the resonance width. A classic example of this
is the effect of the nucleon pole on the shape of the $\Delta (1232)$
resonance, first treated by the theory of Chew and Low [15].
In the present mass range, fitting an $s$-dependent width to each resonance
is an impractical luxury, because there are too many unknowns. For
example, form factors and the opening of new channels will distort the
shape of each resonance.
For simplicity, we adopt constant widths.

One of our objectives is to examine the SU(3) content of resonances.
The simplest approach would be to require each resonance
to have the same phase $\phi _i$ for all its decay channels.
Secondly,
amplitudes for decay to $\eta \eta$ and $\eta \eta '$ are in principle
related to those for $\pi \pi$ via the quark content of the resonance,
and via the well-known composition of $\eta $ and $\eta '$ in terms of
singlet and octet states and the pseudoscalar mixing angle $\Theta $ :
\begin{eqnarray}
|q\bar q > &=& \cos \Theta |\eta >+ \sin \Theta |\eta '> \\
|s\bar s > &=&-\sin \Theta |\eta > + \cos \Theta |\eta '>,
\end{eqnarray}
where $\cos \Theta \simeq 0.8$ and $\sin \Theta \simeq 0.6$ [13].
Amplitudes for decay of $I=0$ $q\bar q$ combinations to $\pi ^0 \pi ^0$,
$\eta \eta$ and $\eta \eta '$ may then be written compactly as:
\begin{eqnarray}
f(q\bar q) &=& \frac {|\pi ^0 \pi ^0 >}{\sqrt{2}} + \frac {\cos ^2 \Theta
}{\sqrt {2}}|\eta \eta > +  \cos \Theta \sin \Theta |\eta \eta '>, \\
f(s\bar s) &=& \sqrt\lambda [\sin ^2 \Theta |\eta \eta > +
\cos ^2 \Theta |\eta \eta '> - \sqrt{2}\cos \Theta \sin \Theta |\eta \eta '>].
\end{eqnarray}
We have introduced into the latter equation a factor $\lambda$. This factor has
been fitted empirically in Ref. [16] to a large variety of data on strange
and non-strange
final states at high energies. Its value is
$\sqrt\lambda = 0.8-0.9$ and we adopt
the central value of 0.85.

The resonances $R$ we observe in present data may be
linear combinations of $q\bar q$ and $s\bar s$:
\begin {equation}
R = \cos \Phi |q\bar q > + \sin \Phi |s\bar s >.
\end {equation}
We use $q\bar q$ as a shorthand to denote $(u\bar u + d\bar d)/\sqrt {2}$.
We expect the observed resonances
to be dominantly $q\bar q$ in view of their production in $\bar pp$ reactions;
however, we would like to test this against the observed branching ratios to
$\pi \pi$, $\eta \eta $ and $\eta \eta '$.
Amplitudes for the decay of $R$ to these three channels are given by the
three equations:
\begin {eqnarray}
f(\pi ^0 \pi ^0) &=& \frac {\cos \Phi}{\sqrt {2}} \\
f(\eta \eta) &=& \frac {\cos \Phi}{\sqrt {2}}(\cos ^2 \Theta +
\sqrt{2\lambda}
\sin ^2 \Theta \tan \Phi ) \\
f(\eta \eta ') &=& \frac {\cos \Phi}{\sqrt {2}}\cos \Theta \sin \Theta
(1 - \sqrt{2\lambda}\tan \Phi ).
\end{eqnarray}

In our earliest attempts to fit data on $\eta \eta$ and $\eta \eta '$,
it immediately became obvious that there are major problems in applying these
SU(3) relations. The best fit we shall show later has a $\chi ^2$ of 10585.
If one sets the phase angles $\phi _i$ of each resonance to the same value
in all three decay channels $\pi ^0 \pi ^0$, $\eta \eta$ and $\eta \eta '$
and takes the resonances to be pure $q\bar q$, $\chi ^2$ increases
dramatically to about 45000. The fit is visibly awful.
Some deviation from strict $SU(3)$ is
obviously required. It is necessary to introduce some latitude into
phase angles $\phi _i$ to different channels, or into the flavour
composition of the resonances, or both.

To examine where the problem lies, we have tried fitting two extreme scenarios.
The first is to take all resonances to be pure $q\bar q$ by setting all
$\Phi$ to zero; if the phases $\phi _i$ of all resonances are left free,
$\chi ^2 \to 14558$. This is not a huge increase in $\chi ^2$ over our best
value of 10585.
However, the fit to both $\eta \eta$ and $\pi ^0\pi ^0$ angular
distributions is visibly poor, particularly for $\eta \eta$. Furthermore,
some of the phases $\phi _i$ depart from zero by unreasonably large amounts
$>90 ^{\circ }$.
So it seems that some departure from strict SU(3) is unavoidable.

The other extreme is to set all $\phi _i$ to zero and attempt to fit purely
by adjusting flavour mixing angles $\Phi$. This turns out to be more
successful, giving a $\chi ^2$ of 11527. We shall give detailed results below.
Most of the mixing angles optimise in the range 0--20$^{\circ }$, i.e.
close to pure $q\bar q$ states.
These small mixing angles allow modest changes in branching ratios
between $\pi ^0 \pi ^0$, $\eta \eta$ and $\eta \eta '$.
However, one resonance, $f_0(2105)$,
requires considerably larger mixing, with $\Phi$ in the range
59--71.6$^{\circ }$.
The integrated cross sections shown below for $\pi ^0 \pi ^0$, $\eta \eta$ and
$\eta \eta '$ are fairly close to the values predicted by simple SU(3) on
average.
The question therefore is whether it is wise to set phase angles $\phi _i$
strictly at zero.
We now consider several considerations bearing on this point.

Firstly, each resonance rides on the tails of other resonances.
We have had experience
of fitting data at lower masses by both T-matrix techniques [17] and the
K-matrix [18].
From this experience, we have learned that mixing between
states commonly gives rise to deviations of phase angles $\phi _i$ betweeen
different decay channels in the range --15 to +15$^{\circ }$.
Even if K-matrix poles have the same phases, T-matrix poles can have different
phases; the difference depends on the separation of poles, resonances
widths and the mixing through decay channels.
In view of this experience, we feel it unwise to demand strictly the same
phases $\phi _i$ in all channels.
For most resonances, we have allowed $\phi _i$ to optimise freely for each
decay channel with differences limited
to the range --15 to +15$^{\circ}$.
In isolated cases, where they seem to need it, this
range has been allowed to extend to $\pm 30 ^{\circ}$.

Angles $\Phi$ for flavor mixing mostly optimise naturally in the range --15
to +15 $^{\circ }$. However, for large spin, it is likely that differences in
centrifugal barriers and form factors between $\pi \pi$, $\eta \eta$ and
$\eta \eta '$ may affect branching ratios.
Resonances are commonly believed to have radii of 0.8--1.0 fm.
In $\pi \pi$ decays, the wavelength of each outgoing $\pi$ is 1.1 fm for a
resonance mass of 2.2 GeV.
Overlap of wave functions will play a strong role in determining matrix
elements.
Even quite small momentum differences between $\pi \pi$ and $\eta \eta$
and larger differences for $\eta \eta '$ may lead to significant departures
from SU(3).
We observe a problem in fitting $\eta \eta$ cross sections simultaneously with
$\pi ^0 \pi ^0$ unless some departure from strict SU(3) is allowed.
Our final compromise is therefore to allow flavour mixing angles
to vary in the range 0 to $\pm 30^{\circ}$.
All optimise naturally in this range, except for $f_0(2105)$, which must
be allowed complete freedom.
As far as resonance
masses and widths are concerned, the freedom in flavour mixing angles
fortunately has little effect; masses and widths are determined by the
locations of singularities, which
are independent of mixing and corresponding branching ratios.

A further unknown is the radius to be given to centrifugal barriers.
The masses fitted to $4^+$ states are sensitive to this radius,
since the centrifugal barrier in the $\bar pp$ channel is strong.
If the $f_4(2020)$ is to be fitted to the PDG value of 2044 MeV, the
radius required is unreasonably large, namely 1.5 fm.
The fitted value is 0.88 fm.

Ratios of amplitudes for $L = J \pm 1$, i.e. $r_J = |f_{J+1}|/|f_{J-1}|$
are fitted to real constants for each resonance.
Since this linear combination refers to mixing within the
$\bar pp$ channel, it is taken to be the same for decays to $\pi \pi$,
$\eta \eta$ and $\eta \eta '$.
For the $f_4(2020)$, one expects the channel with
$L = J-1$ to dominate over $L = J + 1$, because of the centrifugal barrier.
This is precisely what we find. Likewise, the $5^-$ state is dominantly
$^3G_5$. This mixing is determined very precisely by the polarisation data.
For $6^+$, we assume pure $^3H_6$.

The data on $\pi ^0 \pi ^0$, $\eta \eta$ and $\eta \eta '$ stop at 1940 MeV/c.
Polarisation data stop at 2200 MeV/c.
From $\pi ^-\pi ^+$ data above 2 GeV/c, it is clear that a $6^+$ resonance is
required at $2485 \pm 40$ MeV.
A resonance close to this mass has been observed decaying to $\pi \pi$
by the GAMS collaboration [19], at 2510 MeV.
It seems likely that there will be further high spin resonances of similar
mass. They lie so close to the top of the available mass range that we cannot
establish their parameters precisely. Nonetheless, it is convenient to use
$5^-$, $4^+$,
and $2^+$ resonances in the range 2500--2620 MeV to parametrise the
data at the highest momenta. The tail of the $6^+$ resonance in the
mass range LEAR below 2410 MeV needs to be described accurately, since its
interferences with states of lower spin are important for fitting
Legendre polynomials up to order 10 required by the $\pi \pi$ and
$\eta \eta$ data.

\section {Results}
We summarise first the expected states.
The $f_4(2020)$ is well known and there is plenty of evidence, summarised
by the Particle Data Group, for its radial excitation $f_4(2300)$.
There is a known $\rho _5$ at 2330 MeV.
Consequently, resonances having spins with all $J^P$ below $4^+$
are expected around 2020 MeV and with all $J^P$ below $5^-$ around
2300 MeV.
In the mass range 1910--2410 MeV, our main focus of attention,
this implies two $^3D_3$ $q\bar q$ states plus one with $^3G_3$ near
2300 MeV, two $^3P_2$ and two $^3F_2$ states, four $1^{--}$ states and
two $0^+$.
We are able to identify all of these except for two $1^-$ states.
It is hardly surprising that it is difficult to disentangle two states
of low $J^P$.

\subsection {Uniqueness}
The ambiguities we have encountered concern signs of the amplitude
ratios $r_4 = ^3H_4/^3F_4$ and $r_2 = ^3F_2/^3P_2$.
These are the ratios of amplitudes after centrifugal barrier factors are
factored out, i.e. they are asymptotic ratios as $s \to \infty$.
We have located two solutions with similar $\chi ^2$, one with
$r_4$ positive and the other with $r_4$ negative.
They both contain the same resonances with similar masses and widths,
and differ mostly in coupling constants.
However, analysis of $\eta \pi ^0 \pi ^0$ data [12,20] allows a clear
distinction between these alternatives.
Clebsch-Gordan coefficients in equns. (4) and (5)
are such that positive $r_4$ requires
dominance of $M = 1$ amplitudes for $f_4(2300)$.
[Here $M$ is the projection of the spin of the initial state along the
beam direction; $M = 1$ corresponds to $f_{+-}$ in the helicity basis].
This agrees with the $\eta \pi ^0 \pi ^0$
analysis. Negative $r_4$ is rejected strongly by that analysis.

There is an argument concerning matrix elements for decay which supports this
conclusion.
For this high spin, the wave function of the resonance
is peaked strongly at a radius of 0.8--1.0 fm.
Wave functions for the $\bar pp$ channel are described by spherical
Bessel functions.
For momenta in the mass range under
discussion, the first zero of $j_L(kr)$ lies
outside 2 fm for both $L = 3$ and $L = 5$.
Overlap of wave functions with the resonance leads to the
expectation that $\bar pp$ $^3F_4$ and $^3H_4$ will
couple to the resonance with the same sign, hence $r_4$ positive.
We find that the fitted amplitude ratios $r_5 = ^3I_5/^3G_5$ and
$r_3 = ^3G_3/^3D_3$ are likewise positive, as one would expect from
the same argument.

We shall present evidence for four $2^+$ resonances at 1910, 2020, 2230
and 2300 MeV. We shall argue for several reasons that the $f_2(2020)$ is the
first $q\bar q$ $^3F_2$ state. We have examined all alternative sign
combinations of $r_2$ for these four resonances.
The $\pi ^- \pi ^+$ polarisation data are sensitive to these signs.
The best solution is obtained with signs respectively +, +, -, - in the order
$f_2(1910)$ to $f_2(2300)$.
The second best solution has signs -, -, -, -, but a $\chi ^2 $ of 13850,
i.e. worse by 3265 than the best solution.
The increase in $\chi ^2$ is mostly for polarisation data, where
there is a clear systematic discrepancy between data and fit for the
second alternative.
For $f_2(2020)$, $r_2$ is again expected to be positive.
The first zero of the Bessel function
$j_1(kr)$ (coupling to $^3P_2$) again lies outside 2 fm;
so the $\bar pp$ momentum is low
enough that the argument given above remains unchanged.

For $f_2(2300)$, which we shall interpret as the radial excitation of
$f_2(2020)$, the momentum in the $\bar pp$ channel is such that the
first zero of $j_1(kr)$ lies at 1.28 fm.
When one allows for (i) the fact that the wave function will be attracted to
smaller $r$ by the resonant interaction (mixing in a component due to
the irregular function $n_1(kr)$), and (ii) the node
in the radial wave function of the resonance, it is possible to
arrive at a negative $r_2$, as observed.
For $q\bar q$ $^3P_2$ states, it is not possible to make reliable
predictions, because of the large number of nodes in the radial wave functions
of the resonances. They are, however, found to show the same general
variation with $s$ as for other $J^P$, i.e. positive for low $\bar p$
momenta and going negative for high momenta.

\subsection {Quality of the fit}
A solution is obtained in typically 1-2 minutes of computing;
consequently several thousand  fits have been made, exploring
the systematic effects of including or omitting various resonances.
In the course of this work, we have found, from the
observed variations from fit to fit, that
errors on resonance masses and widths may be assigned from
$\chi ^2$ changes of $\sim 20$. This is the quantitative criterion we
adopt in assessing errors given here.
It is larger than statistics ($\Delta \chi ^2 = 1)$ and therefore conservative.
It covers almost
all of the variations which have been observed between fits
with differing ingredients; where it does not, errors have
been increased to cover the observed variations.

The $\pi ^0 \pi ^0$ data have statistical errors which are much smaller
than  those of other data.
Consequently, there is the danger that delicate features of the $\pi ^0
\pi ^0$ data cause problems in fitting other channels.
To remedy this, we reduce the weight of the $\pi ^0 \pi ^0$ data by
a factor 3, and increase the weight of $\eta \eta '$ by a factor 2.
This does not introduce any major qualitative changes, but speeds
convergence greatly.

Values of $\chi ^2$ for all data sets are given in Table 1. They are somewhat
above 1 per point, particularly for $\pi ^0 \pi ^0$, because
of the weighting mentioned above. It may reflect the presence of systematic
errors in some data points, but also may reflect small missing components
in the fit (e.g. backgrounds from the
tails of lower mass resonances), which cannot presently be
identified unambiguously. Our objective is to locate the essential
features of the data, and these are stable against small variants in the
parametrisation of the amplitudes.

\begin{table} [htp]
\begin{center}
\begin{tabular}{cccc}
Data & Author & Points & $\chi^2$\\\hline
$\pi ^-\pi ^+$ & Hasan & 1000 & 2677\\
$P$            & Hasan & 1000 & 1812\\
$\pi ^-\pi ^+$ & Eisenhandler & 960 & 1702\\
$P$            & Eisenhandler & 1078 & 1720\\
$\pi ^0\pi ^0$ & Anisovich    & 360  & 1853 \\
$\eta \eta $   & Anisovich    & 360  & 667 \\
$\eta \eta '$  & Anisovich    & 144  &  142 \\
Normalisations & Anisovich    & 27   &  12 \\\hline
\end{tabular}
\caption {Contributions to $\chi ^2$ of the fit.}
\end{center}
\end{table}

\subsection {Normalisation}
Each of the data sets on $\pi ^0 \pi ^0$, $\eta \eta$ and $\eta \eta '$ is
given a normalisation constant which has been determined in Ref. [1].
In our final fits, the normalisation is optimised by
allowing each data set to vary in normalisation and including into
$\chi ^2$ a contribution for the deviation of the normalisation from its
experimental value.

As described in [1], the normalisation of Crystal Barrel data for
$\pi ^0 \pi ^0$ is on average slightly more than a factor 2 higher than that
of data of Dulude et al. [21]. This calls into question the absolute
normalisation of all the neutral data.
In order to test this, we have
tried scaling the normalisation of all Crystal Barrel data for
$\pi ^0 \pi ^0$, $\eta \eta$ and $\eta \eta '$ down by this factor 2, so
as to agree on average with Dulude et al.
It results in a huge increase in $\chi ^2$ from 10585 to 16893.
The polarisation data are particularly sensitive to this change,
increasing from 3532 to 7940.

The reason for this is straightforward, but important.
Partial waves in the $\pi ^0 \pi ^0$ channel have isospin 0 and even spin.
These amplitudes contributes also to $\pi ^-\pi ^+$ data, but
interfere with $I = 1$ components with odd spin.
This interference gives rise to forward-backward asymmetries in the differential
cross section, depending on real parts of interferences.
Values of $Pd\sigma /d\Omega$ contain interferences depending on
imaginary parts of identical interference terms.
The normalisation of Dulude et al. requires much larger $I = 1$ amplitudes
than our data.
A fit to the data of Dulude et al.
alters the phases of interferences, in order to fit asymmetries
in differential cross sections. But these changes are inconsistent with
the polarisation data.
This inconsistency rules out the normalisation of Dulude et al.
If we float the normalisation of Crystal Barrel data freely, it
optimises at 0.989 times the published values. An increase of
$\chi ^2$ of 20 (the criterion we have adopted above in discussing resonance
masses and widths) corresponds to a change of $\pm 0.023$ in normalisation.
This lies within the normalisation errors quoted for Crystal Barrel
data, namely $\pm 0.03$ for momenta of 1050 MeV/c upwards and $\pm 0.06$
at 600 and 900 MeV/c.

In the final analysis, we do not use the factor 0.989.
The normalisation of Crystal Barrel data is used
with its experimental normalisation and with the errors (3--6\%)
obtained experimentally.

We include the data of Dulude et al. for the determination of the shape of the
angular distribution. They provide some valuable
points close to $\cos \theta = 1$. However, their normalisation is
fitted freely.

\begin{figure}
\begin{center}
\vskip -25mm
\epsfig{file=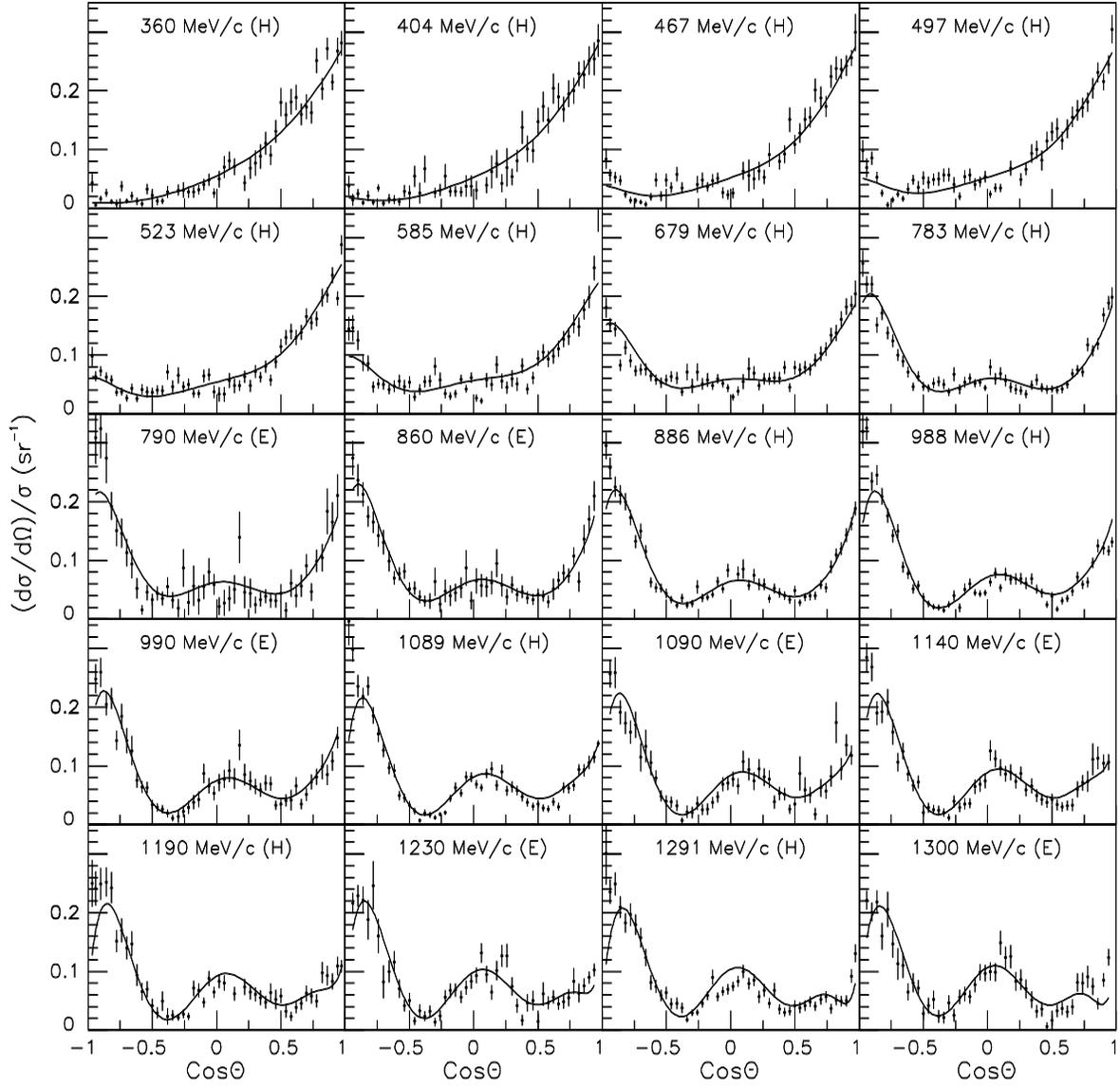,width=18cm}
\vskip -180.47mm
\epsfig{file=PPCH_A.PS,width=18.cm}
~\
\vskip -6mm
\caption{Differential cross sections from 360 to 1300 MeV/c, compared with the
fit (full curves); panels labelled H are data of Hasan et al. and those
labelled E are from Eisenhandler et al. Each distribution is normalised so as
to integrate to $2\pi$.}
\end{center}
\end{figure}
\subsection {Comparison of data and fit}
Figs. 1--5 show comparisons of our final fit with data for $\pi ^- \pi ^+$
differential
cross sections, polarisation and integrated cross sections. The
$\pi ^0 \pi ^0$ data, $\eta \eta$ and $\eta \eta '$ are available only
up to centre of mass scattering angle $\theta$ given by $\cos \theta  \le
0.875$. Their integrated cross sections are therefore compared in Fig. 5
with the fit
only over this angular range. The $\pi ^-\pi ^+$ data are available over the
full angular range and are integrated over this range.

\begin{figure}
\begin{center}
\vskip -10mm
\epsfig{file=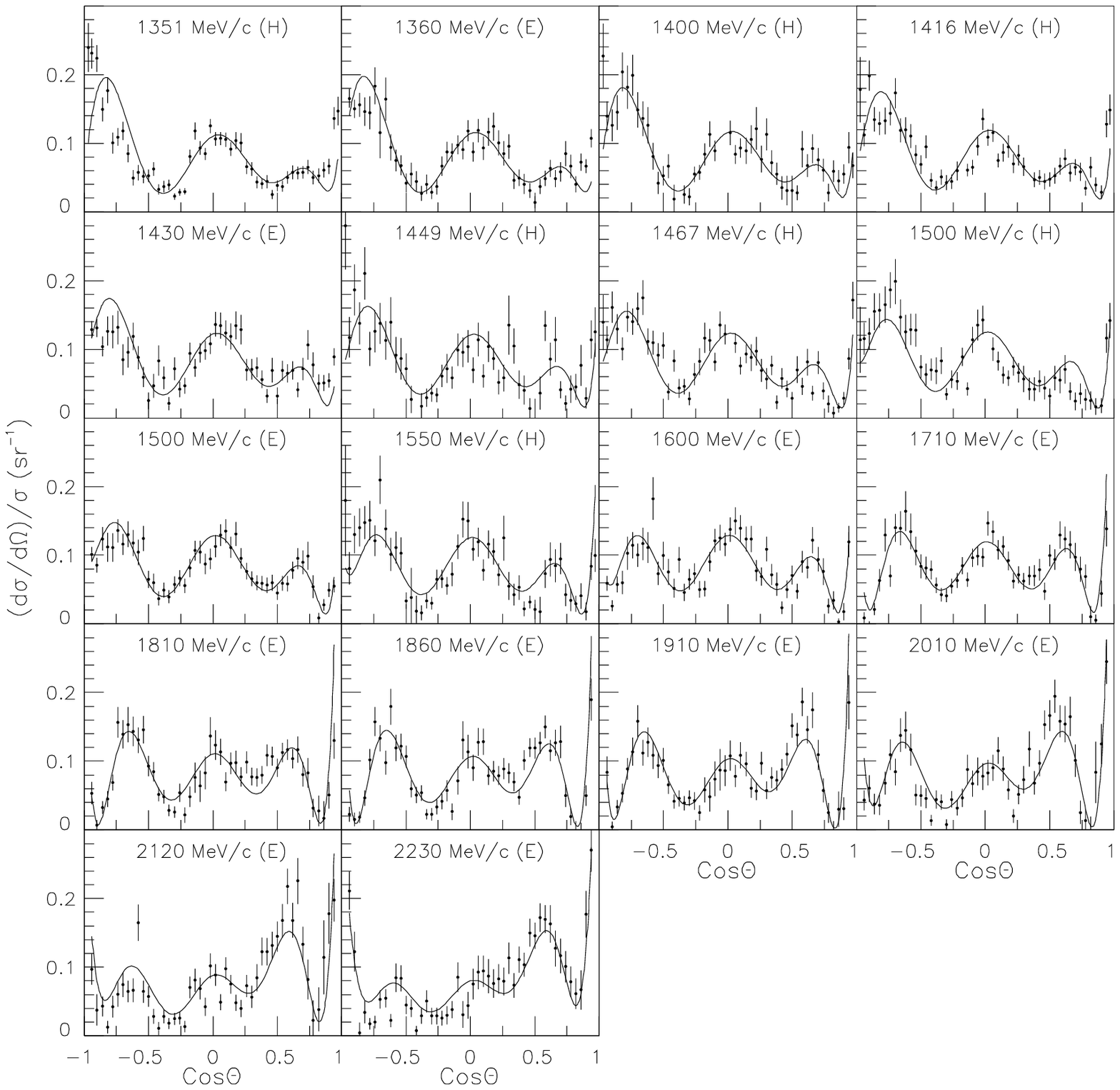,width=18cm}
\vskip -180.47mm
\epsfig{file=PPCH_B.PS,width=18cm}
~\
\vskip -6mm
\caption{As Fig. 1, 1351 to 2230 MeV/c.}
\end{center}
\end{figure}
Panels on Figs. 1--3 labelled H
refer to Hasan et al; those labelled E refer to Eisenhandler et al. or the same
experiment of Carter et al.
On Fig. 1, the angular distributions are mostly
fitted well. There are small discrepancies near $\cos \theta$ = -1 for
momenta of 790, 988, 1089, 1140,  and 1291 MeV/c.
However, in all cases, data from the other
experiment at nearby momenta show either no discrepancy or small ones.
We therefore take these discrepancies to be due to experimental error
or statistics.
There is some discrepancy with $d\sigma /d\Omega$ near $\cos \theta$ = -1
at 467,497 and 585 MeV.
We can find no solution which resolves this discrepancy without rapid
changes with mass in high partial waves $J = 4$, 5 or 6; at these momenta,
contributions from $J = 5$ and 6 should be small.

On Fig. 2 there are some discrepancies at individual momenta, but nothing
systematic over a range covered by a resonance.
Above 2 GeV/c, we are fitting only
with tails of resonances from lower masses, plus resonances with $J ^P = 6^+$,
$5^-$, $4^+$, and $2^+$ in the range 2500--2600 MeV.
It seems likely that there
will be further low spin resonances in this mass range. We have tried adding
one by one resonances with $J^P = 3^-$, $1^-$ and $0^+$, but none gives a
significant improvement in $\chi ^2$.
For polarisation data on Fig. 3, the agreement between data and fit is
generally satisfactory and free of systematic trends.

\begin{figure}
\begin{center}
\vskip -10mm
\epsfig{file=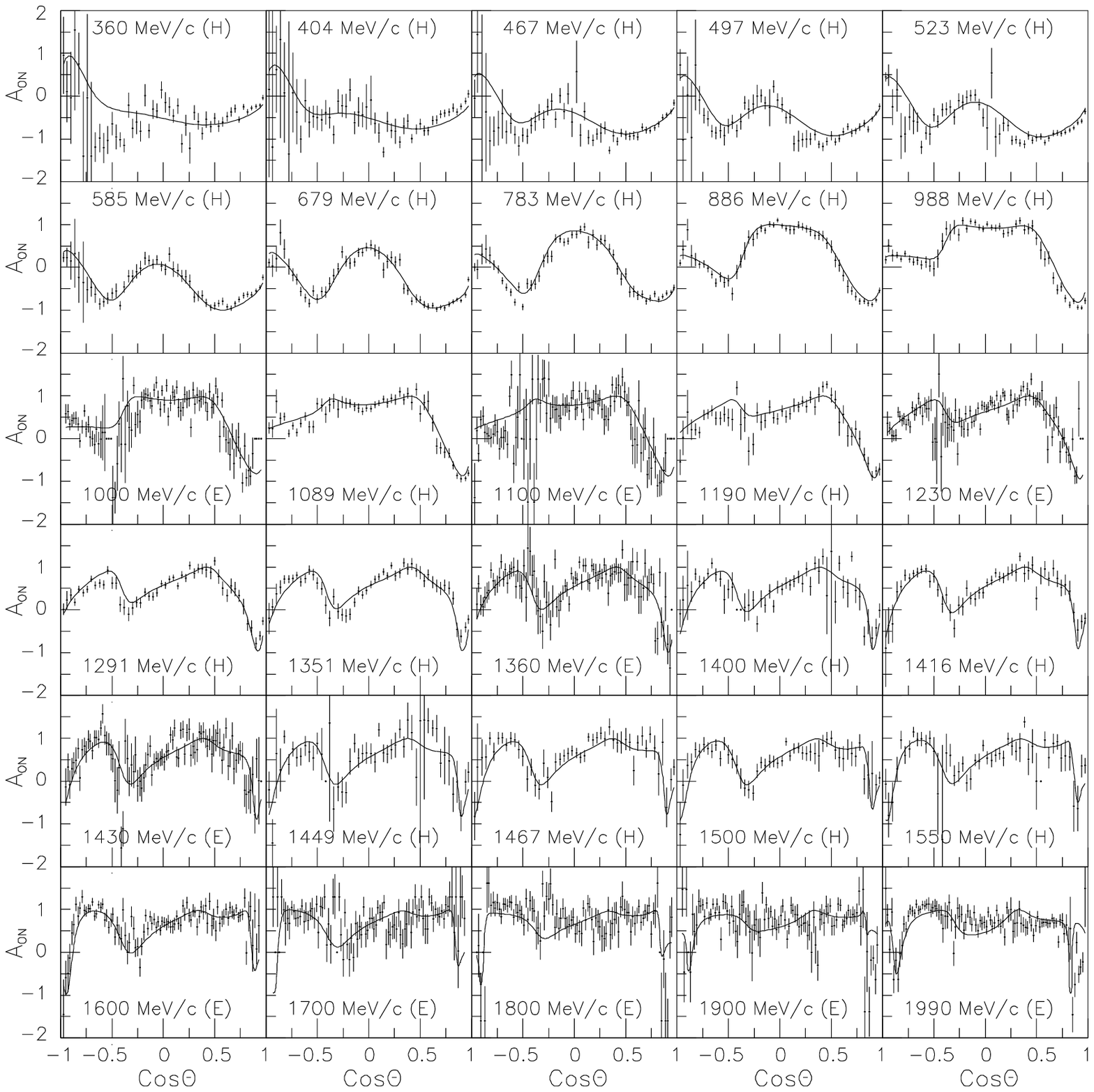,width=18cm}
\vskip -180.47mm
\epsfig{file=POLAR.PS,width=18cm}
~\
\vskip -6mm
\caption{Polarisation data compared with the fit (full curves).}
\end{center}
\end{figure}

\begin{figure}
\begin{center}
\vskip -10mm
\epsfig{file=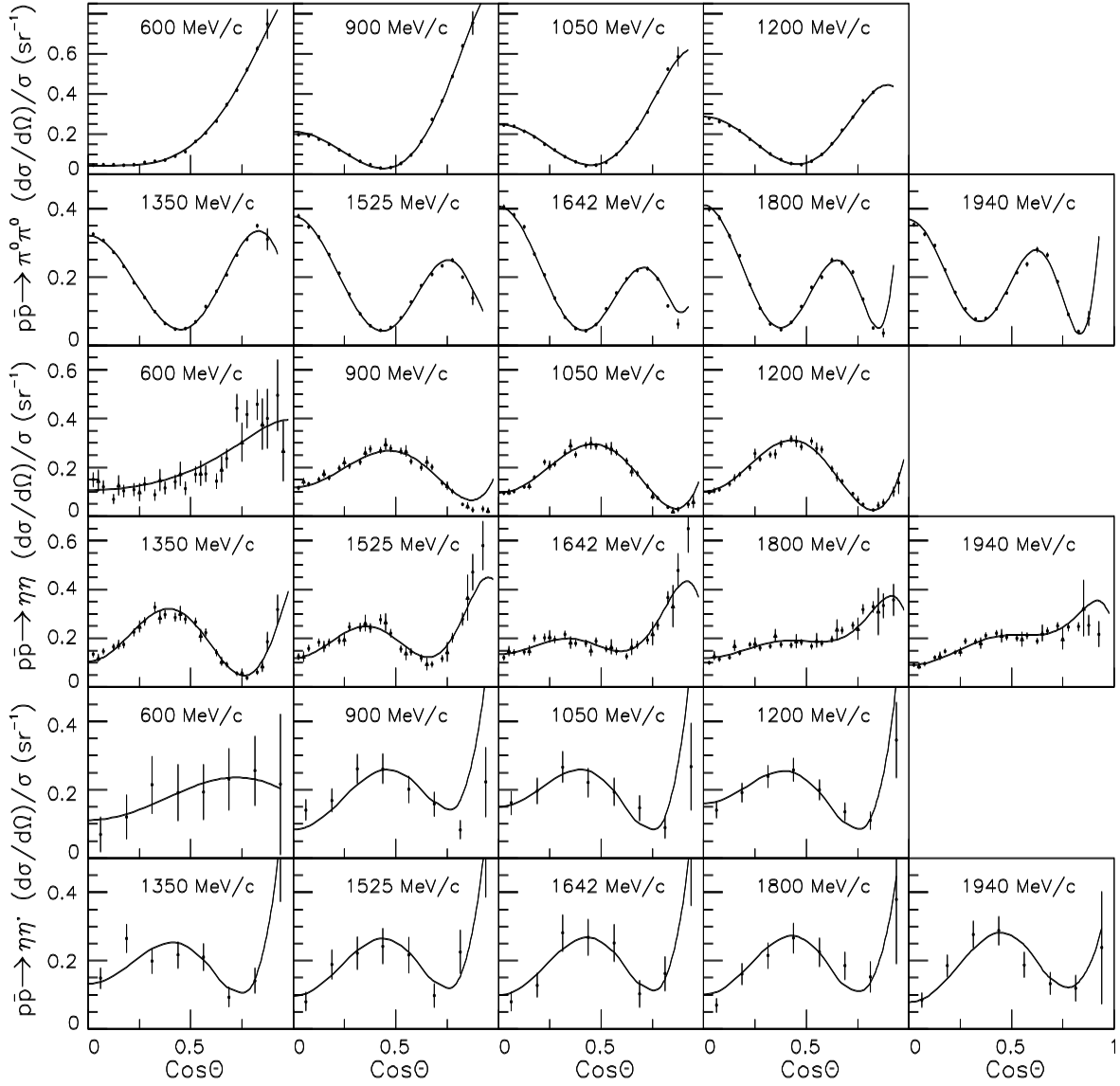,width=18cm}
\vskip -180.47mm
\epsfig{file=NEUT1.PS,width=18cm}
~\
\vskip -6mm
\caption{Differential cross sections for $\pi ^0 \pi ^0$, $\eta \eta$ and
$\eta \eta '$ compared with the
fit (full curves). For $\eta \eta '$, these are the average of $4\gamma$ and
$8\gamma$ data reported in Ref. [1].}
\end{center}
\end{figure}

The fit to $\pi ^0 \pi ^0$ data is shown in Fig. 4 and the accompanying paper
on data analysis [1]. It is good at most momenta.
At 1200 MeV/c, one point at $\cos \theta = 0.875$ cannot be fitted
under any circumstances. The acceptance of the detector falls rapidly around
$\cos \theta = 0.875$, so we suspect this point is affected by some
systematic error.
The data at 1200 MeV/c were actually taken in a run separate from all other
momenta, possibly accounting for a discrepancy between this momentum and
others.

For $\eta \eta$ data, shown on Fig. 4 and in the accompanying paper,
the fit is generally good, though
there are some discrepancies near $\cos \theta = 1$ at 900 and 1940
MeV/c.
Fits to these regions may be improved by relaxing the constraints on
phase angles $\phi _i$.
At 600 MeV/c, statistics are
low and it is possible to obtain some variety of fits, varying the
strengths of flavour mixing for $f_4(2020)$, $f_2(1910)$ and $f_2(2020)$.
Near $\cos \theta = 1$, there is a dip in $\eta \eta$ differential cross
sections at 900 MeV, but a peak at 600 MeV/c. Our fits are a compromise
between these rapidly changing features. The fit to either momentum may be
improved, but not both simultaneously.
For $\eta \eta '$, fits are satisfactory, but statistics are low.

Integrated cross sections, shown on Fig. 5, are described  well.
For $\pi ^0\pi ^0$ in Fig. 5(b), the peak at 900 MeV/c is mostly
due to $f_4(2020)$.
For $\eta \eta$ in Fig. 5(c),
the striking peak at 1200 MeV/c comes largely from
$f_0(2105)$ and $f_4(2020)$.

\begin{figure}
\begin{center}
\vskip -16mm
\epsfig{file=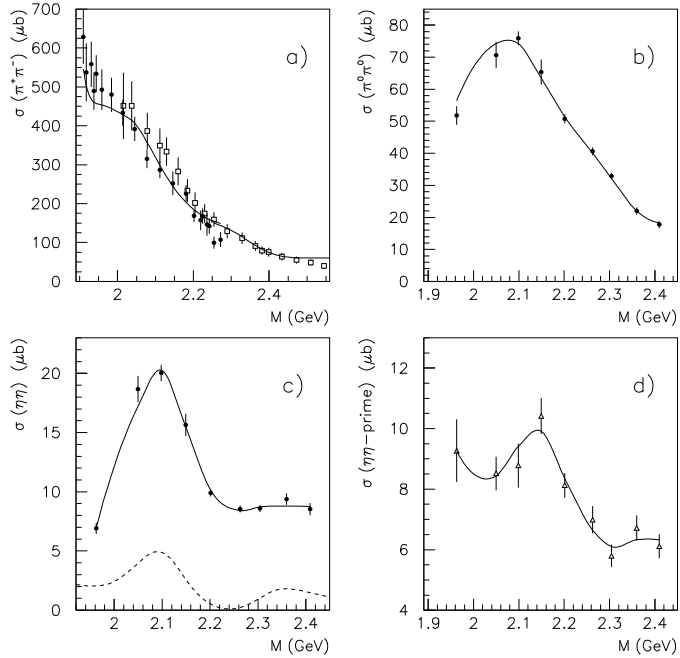,width=10cm}
~\
\vskip -5mm
\caption{Integrated cross sections compared with the fit (full curves). For
$\pi ^0 \pi ^0$, $\eta \eta$ and $\eta \eta '$, the integration is from $\cos
\theta$ = 0 to 0.875 and is over only one hemisphere;
the $\eta \eta '$ data are averaged over $4\gamma$ and $8\gamma$ data reported
in
Ref. 1. For $\pi ^- \pi ^+$ the integration is
over the full angular range.
In (a), black circles are data of Hasan et al. and open squares are those of
Eisenhandler et al. The dashed curve in (a) shows the contribution from
the $0^+$ intensity.}
\end{center}
\end{figure}

\begin{table} [htp]
\begin{center}
\begin{tabular}{ccccccc}
$J^P$ & Mass $M$ & Width $\Gamma$  & $\Delta \chi ^2$
& Hasan $M$ & Hasan $\Gamma $ \\
& (Mev) & (MeV) &  & (MeV) & (MeV) \\\hline
$6^+$ & $2485 \pm 40$ & $410 \pm 90$  & 1776 & - & - \\
$5^-$ & $\sim 2500$ & $\sim 470$ &112  & 2881 & 310 \\
$5^-$ & $2295 \pm 30$ & $235 ^{+65}_{-40}$  & 2534 & 2303 & 169 \\
$4^+$ & $\sim 2500$ & $\sim 400$ &1305  & 2813 & 257 \\
$4^+$ & $2300 \pm 25$ & $270 \pm 50$ &2549 & 2314 & 278 \\
$4^+$ & $2020 \pm 12$ & $170 \pm 15$ &22382 & (2049) & (203) \\
$3^-$ & $2300 ^{+50}_{-80}$ & $340 \pm 150$  & 183 & - & -\\
$3^-$ & $2210 \pm 40$ & $360 \pm 55$ &  368 & 2232 & 220 \\
$3^-$ & $1960 \pm 15$ & $150 \pm 25$ &2957  & 2007 & 287\\
$2^+$ & $\sim 2620$ & $\sim 430$ & 776  & - & -\\
$2^+$ & $2300 \pm 35$ & $290 \pm 50$ & 2879  & 2517 & 264\\ \
$2^+$ & $2230 \pm 30$ & $245 \pm 45$ & 2290 & 2226 & 226 \\
$2^+$ & $2020 \pm 30$ & $275 \pm 35$ & 2980  & 1996 & 134 \\
$2^+$ & $1910 \pm 30$ & $260 \pm 40$ & 2286  & - & - \\
$1^-$ & $2165 \pm 40$ & $160 ^{+140}_{-70}$  & 450 & 2191 & 296\\ \
$1^-$ & $2005 \pm 40$ & $275 \pm 75$ & 1341  & 1988 & 244\\
$1^-$ & $(1700)$ & ($180$) &8444  &  (1690) & 246 \\
$0^+$ & $2320 \pm 30$ & $175 \pm 45$ &1257  & 2321 & 223\\ \
$0^+$ & $2105 \pm 15$ & $200 \pm 25$  & 4030 & 2122 & 273 \\
$0^+$ & $2005 \pm 30$ & $305 \pm 50$  & 370 & - & -\\
$0^+$ & $(1700)$ & $1000$ &  2844 & 1745 & 238 \\\hline
\end{tabular}
\caption {Resonances fitted to the data; errors cover the full range of
systematic variations
observed in a large variety of fits with varying ingredients. Values in
parentheses either lie outside the range of
masses fitted and describe background amplitudes or are fixed.
The fourth column gives
changes in $\chi ^2$ when each resonance is removed from the fit and all other
parameters are re-optimised. The final two columns compare
with results of Hasan and Bugg, Ref. [9]. }
\end{center}
\end{table}
\subsection {Fitted Resonances}
Fitted resonances are shown in Table 2. We shall comment in detail on most
of them. All those in the mass range 1900--2400 MeV are established
securely, with the  exception of the $3^{--}$ resonance at 2300 MeV,
which is rather weak.
The fourth column of the table shows changes in $\chi ^2$ when each resonance is
removed from the fit and all other parameters are re-optimised; this measures
the significance of each contribution in fitting the data.
Most of these changes are enormous and leave no possible doubt about the
presence of these resonances. Our impression, from the
consistency between alternative fits with differing ingredients,
is that any resonance contributing a change in $\chi ^2>
100$ is definitely present. However, below $\Delta \chi ^2 = 300$,
the mass and
particularly the width are hard to establish with confidence.
Those resonances contributing 1000 to the improvement in $\chi ^2$ are
very secure in all parameters.
\begin{figure}
\begin{center}
\vskip -15mm
\epsfig{file=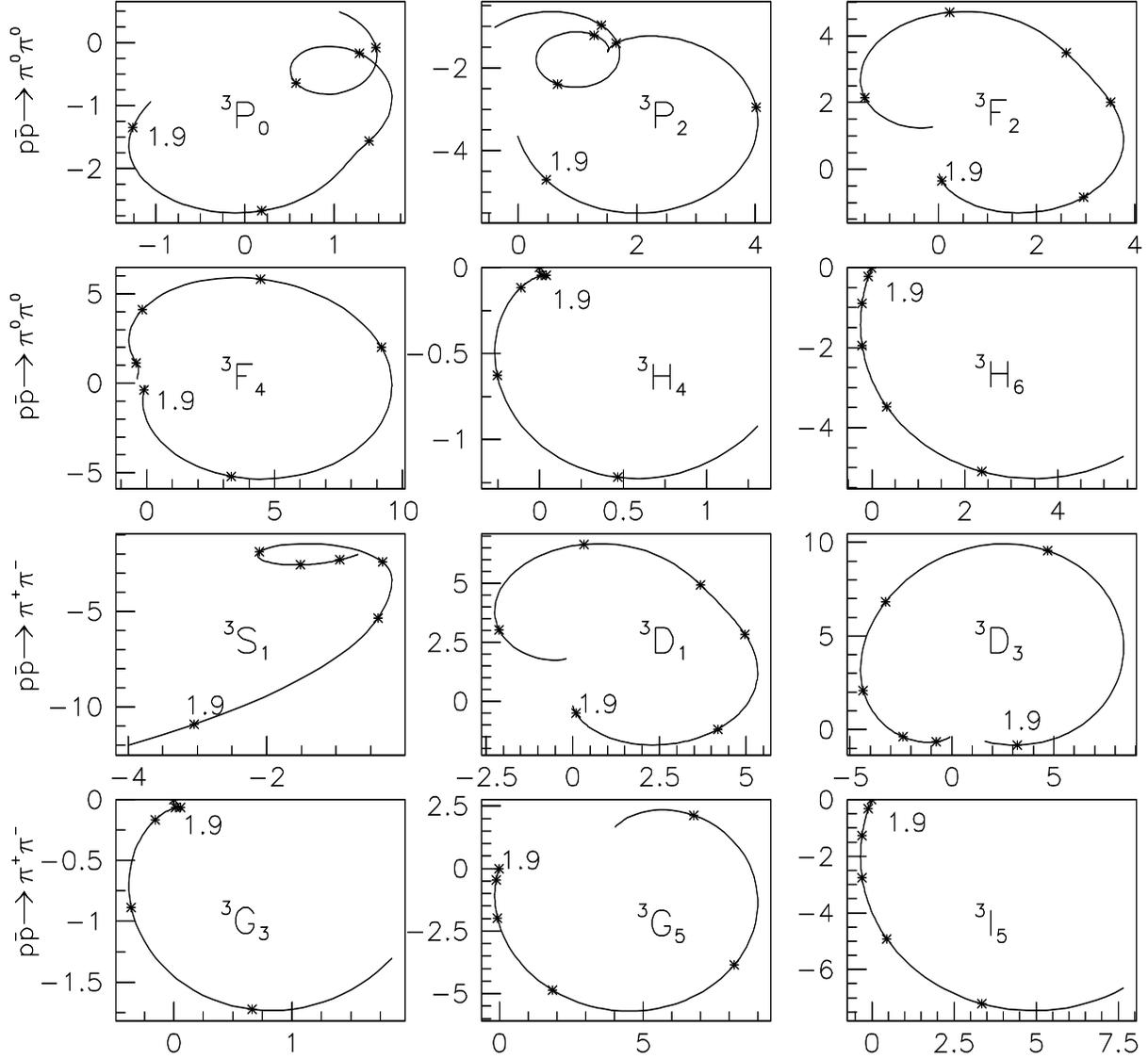,width=18cm}
\vskip -180.45mm
\epsfig{file=ARGANT1.PS,width=18.cm}
~\
\vskip -6mm
\caption{ Argand diagrams for $\pi ^+ \pi ^-$ partial wave amplitudes
$f_J(s)$.
Crosses mark masses at 100 MeV intervals beginning at 1900 MeV.
The scale is such that the integrated partial wave cross section in given
in $\mu b$ by $2\pi |f_J(s)|^2\rho (s)$, where $\rho (s)$ is the
phase space $q/\sqrt {s}$. }
\end{center}
\end{figure}

Argand diagrams are shown for $\pi \pi$ amplitudes on Fig. 6 and
those for $\eta \eta$  and $\eta \eta '$ on Fig. 7. Resonance
loops are observed in all partial waves. We shall comment in detail
later. Intensities of all partial waves are displayed in Fig. 8.
Dashed contributions are for $L = J-1$ and dotted for $L = J+1$. [Interferences
disappear from their sum].

There is a strong $4^+$ resonance at 2020 MeV and a somewhat weaker, but
definite, one at 2300 MeV; there is
tentative evidence for another around 2500 MeV. A strong $5^-$ state is
observed at 2295 and there is tentative evidence for a higher one at
$\sim 2500$ MeV. There is definite evidence for the expected $6^+$
resonance at $2485 \pm 40$ MeV.
For $J^P = 3^-$, there are two well defined $^3D_3$ states at
1960 and 2210 MeV.
There is tentative evidence for a further $^3G_3$ state at 2300 MeV,
but it is weak; despite this, its angular dependence is
distinctive and it is hard to avoid the conclusion that it is
present, even if its mass and width cannot be determined well.

\begin{figure}
\begin{center}
\vskip -15mm
\epsfig{file=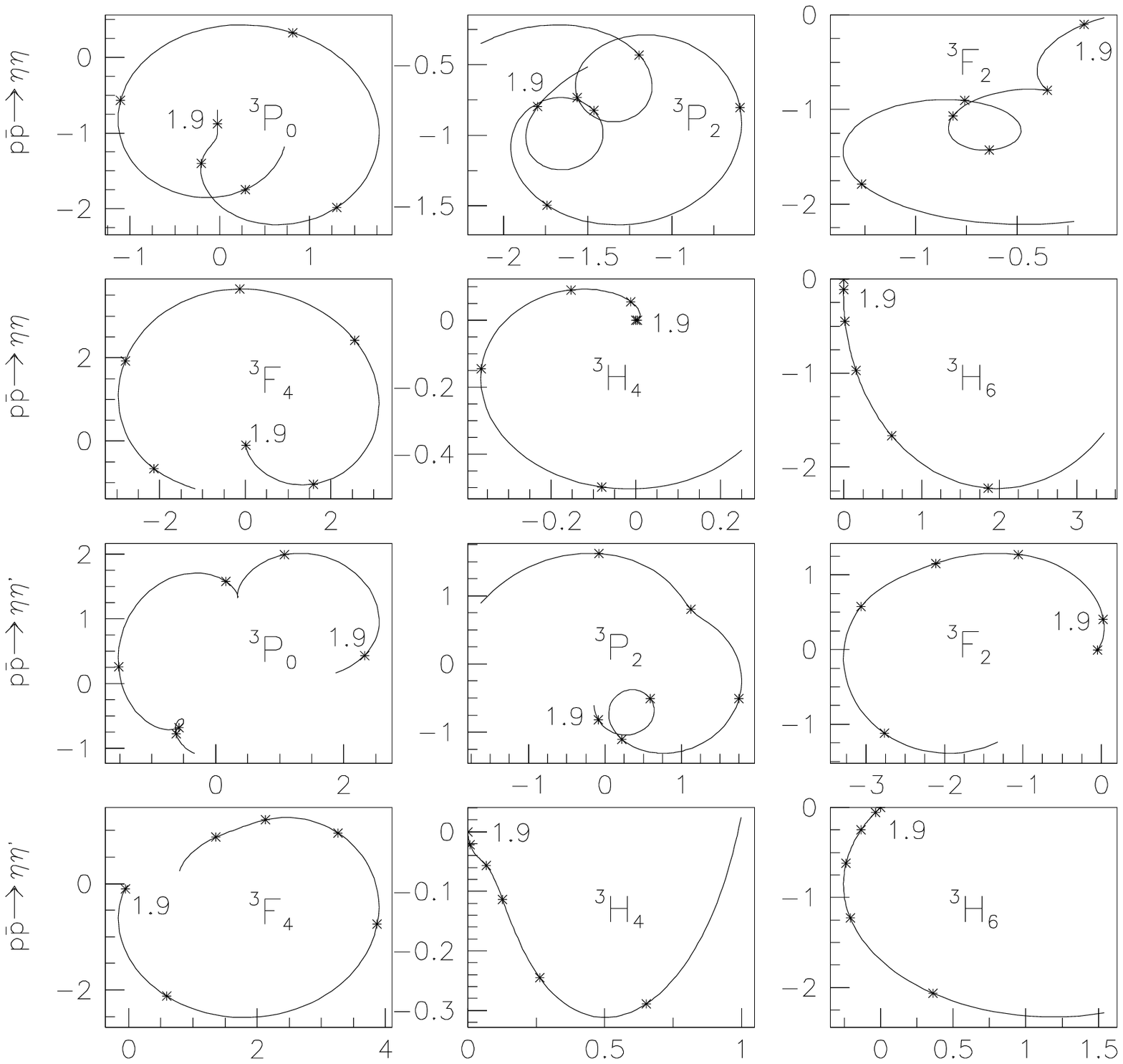,width=18cm}
\vskip -180.46mm
\epsfig{file=ARGANT2.PS,width=18cm}
~\
\vskip -6mm
\caption{Argand diagrams for $\eta \eta$ and $\eta \eta '$ amplitudes.
Crosses mark masses at 100 MeV intervals beginning at 1900 MeV.
The scale is as described for Fig. 6.}
\end{center}
\end{figure}

\begin{figure}
\begin{center}
\vskip -10mm
\epsfig{file=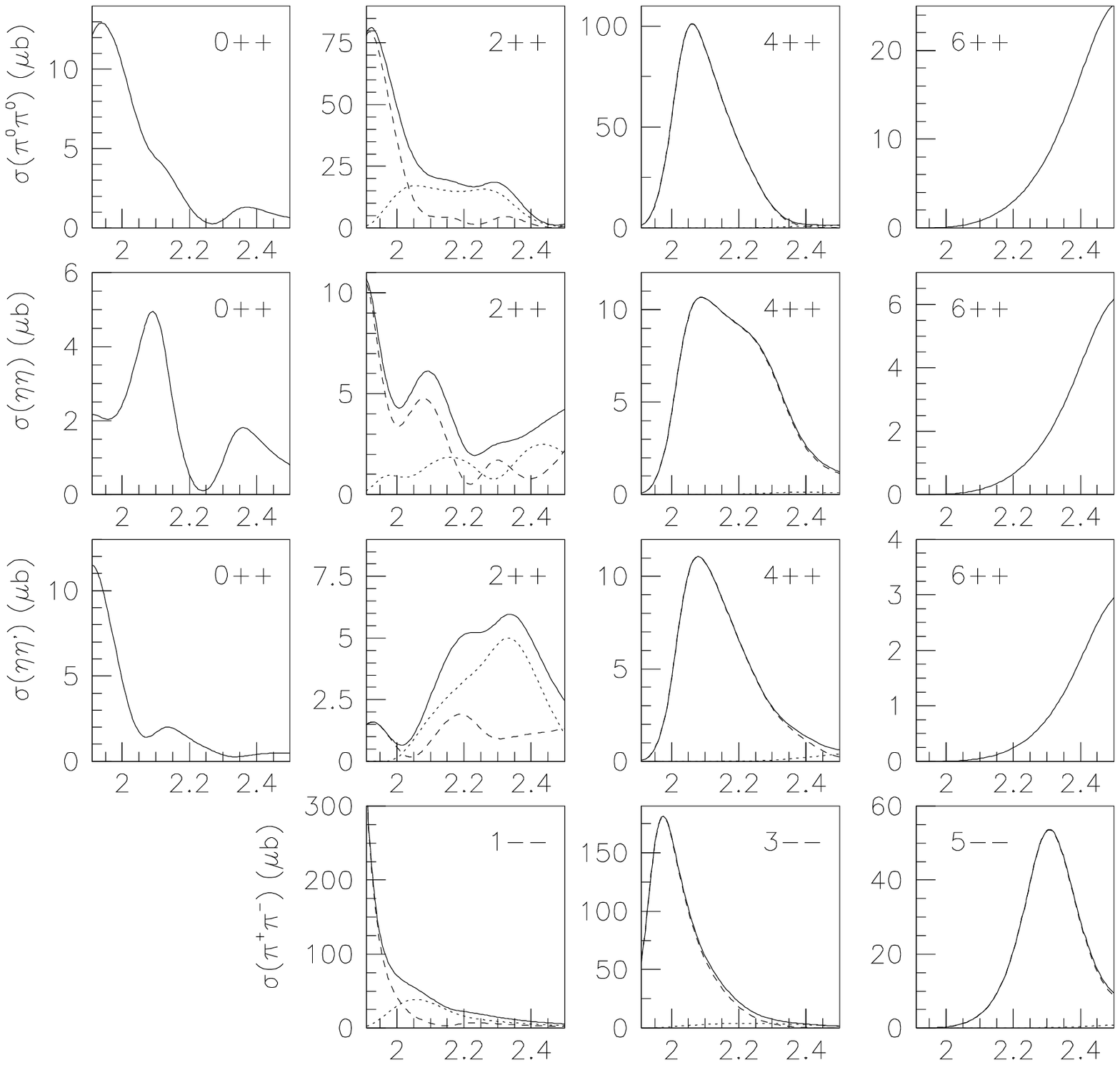,width=18cm}
\vskip -180.46mm
\epsfig{file=PW_ANAL.PS,width=18cm}
~\
\vskip -6mm
\caption{Intensities from individual $J^P$ for channels $\pi ^0 \pi ^0$, $\eta
\eta$ and $\eta \eta '$. Dashed curves show contributions with $J = J-1$ and
dotted curves those with $L = J+1$.}
\end{center}
\end{figure}

An important result is that there are definitely four $2^+$ states
at 1910, 2020, 2230 and 2300 MeV;
these are expected from the quark model for $^3P_2$ and $^3F_2$.
For $J^P = 1^-$, four states are again expected. There is evidence that
two are required, and there are tentative indications of a third;
however, it is not possible to
establish the mass and width of this third state.
For $J^P = 0^+$, at least two resonances are definitely required.
The one at 2105 MeV appears very strongly, particularly in the $\eta \eta$
data, and its mass and width are determined precisely.
There is a further definite $0^+$ resonance at 2320 MeV.
As we shall comment below, there is probably  a third $0^+$ state
at 2005 MeV, but its mass and width are not well
determined because of overlap and interference with $f_0(2105)$.
We now discuss each resonance in turn, starting with the high mass region.

\subsection {The mass range 2500--2600 MeV}
Our main objective is to study the momentum range up to 1940 MeV/c,
where the extensive Crystal Barrel data are available.
However, we also fit the momentum range above 2 GeV/c with the objective of
obtaining a reliable determination of the the tails of the high mass
resonances below 2 GeV/c.
The $\pi ^-\pi ^+$ differential cross section data extend to 2430 MeV/c and
polarisation data to 2200 MeV/c.

We find definite evidence for a $6^+$ resonance at 2485 MeV, in close
agreement with the result of the GAMS collaboration [19].
Its contribution to $\chi ^2$ is large, as one sees from Table 2.
We find it desirable to introduce further high mass contributions for
$5^-$, $4^+$, and $2^+$, as one sees from the changes they
introduce into $\chi ^2$, listed in Table 2.
However, only the lower half of each resonance is within the available
mass range. The result is, of course, a strong correlation between fitted
mass and width, so we are unable to assign accurate parameters to these
resonances.
Furthermore, masses of the high spin states are again sensitive to the
radius of the centrifugal barrier.

We now comment on individual resonances and illustrate on Fig. 9
the deterioration in the fit when they are dropped from the fit.

\subsection {$5^-$ and $4^+$}
The $5^-$ resonance at 2295 MeV is very clear in polarisation data.
Figs. 9 (a)-(c) illustrate the deterioration in fits when it is
removed.
The width of $\rho _5(2295)$ is $235 ^{+65}_{-40}$ MeV, somewhat
less than the 400 MeV quoted by
the GAMS collaboration [22]. However, an inspection of
the GAMS data reveals the possibilty that they have fitted resonances
at 2295 and 2500 MeV by a single resonance.
On Fig. 6 one sees that the $5^-$ state at 2295 MeV is dominantly $^3G_5$
and the well known $f_4(2020)$ is almost purely $^3F_4$.
The centrifugal barriers for
$\bar pp$ will suppress higher $L$ strongly.

The phase advance of 360$^{\circ}$ for $^3F_4$ on Fig. 6
requires the presence of two $4^+$ resonances.
A second one at 2300 MeV is definitely required, but is fairly weak in
the $\pi \pi$ channel, leading to a large uncertainty on its width.
It is more clearly visible in $\eta \eta$ data.
Earlier fits to $\pi \pi$ data
[5--10] all gave masses in the range 2300--2340 MeV.
Crystal Barrel data on the $\eta \pi ^0 \pi ^0 $ final state show a strong
peak in $f_2(1270)\eta$ at $2320 \pm 30$ MeV with $\Gamma = 220 \pm 30$ MeV
[12].
Also recent VES data display conspicuous $4^+$ peaks in $\omega \omega$ at
$2325 \pm 15$ MeV with $\Gamma = 235 \pm 40$ MeV [23] and in $\eta \pi \pi$
data with $M = 2330 \pm 10(stat) \pm 20(syst)$ MeV with
$\Gamma = 235 \pm 20\pm 40$ MeV [24].

\subsection {$3^{--}$ states}
The $3^{-}$ state at 1960 MeV is strong and very secure.
Statistically, the error on the mass is $\pm 7$ MeV and on the width is
$\pm 14$ MeV.
In all fits, it never moves outside the mass range 1950--1980 MeV.
We increase errors to cover all systematic variations.
An interesting
point is that this resonance is definitely lower in mass than
the corresponding $4^+$ states.
This demonstrates that all resonances in this region are not degenerate
in mass.
Again, the phase advance of 360$^{\circ}$ for $^3D_3$ on Fig. 6
requires the presence of at least two resonances.
The higher one agrees well in mass  with the determination
of Hasan and Bugg, namely $M = 2232$ MeV.
Both resonances are coupled dominantly to $\bar pp$ $^3D_3$.

In order to demonstrate that the data really demand the presence of
each of these resonances, we have dropped them one by one from the fit
and examined the discrepancies which emerge between data and fit.
Figs. 9 (d) and (e) illustrate some of the defects when the lower $3^-$
resonance is dropped; Figs. 9(f) and (g) show defects in the fit
when the upper $3^-$ resonance is removed.
In both cases, there is a tendency for the remaining resonance to become
broad and span the mass range.

A further $q\bar q$ $^3G_3$ state is expected in the mass range
around 2300 MeV, close to $^3G_5$.
There is evidence for its presence as $\rho _3(2300)$ of Table 2.
However, it improves $\chi ^2$ only by 183,
with the consequence that its mass and particularly its width are not
well determined.
It appears almost purely in $\bar pp$ $^3G_3$.
The angular dependence of this state is distinct from $^3D_3$, making
its presence very likely, despite the fact that its parameters are
not well determined.
This angular component requires a mass $\sim 100$ MeV above that of
$\rho _3(2210)$.

We have also tried including $\rho _3(1690)$ as a background amplitude.
It has only a very marginal effect, so we omit it. The reason may well
be that it is inhibited near the $\bar pp$ threshold by the $L = 2$
centrifugal barrier.

\subsection {$2^+$ states}
An important outcome of the present analysis is strong evidence for
four $2^+$ states in this mass range, as expected from $q\bar q$ $^3P_2$ and
$^3F_2$.
Earlier, there has been evidence from VES and GAMS groups for
an $f_2(1920)$ of rather narrow width, $\Gamma \simeq 90$ MeV [25,26].
Recent VES data on $\omega \omega$
give an improved mass determination of $1937 \pm 12$ MeV and
width $\Gamma = 150 \pm 17$ MeV [23].
Data on $\eta \pi ^0 \pi ^0$ from Crystal
Barrel [12,20] require $2^+$ states at (a) M$ = 2020 \pm 50$ MeV, $\Gamma =
200 \pm 70$ MeV, decaying dominantly to $f_2(1270)\eta$,
(b) $M = 2240 \pm 40$ MeV with $\Gamma = 170 \pm 50$ MeV, appearing as
a second peak in $f_2(1270)\eta$, and
(c) $M = 2370 \pm 50$ MeV with $\Gamma = 320 \pm 50$ MeV, appearing as
a strong peak in decays to $a_2(1320)\pi$.

The inclusion of polarisation data is important in clarifying the
components required in the fit.
The $f_2(1920)$ appears very strongly with $M = 1910 \pm 30$ MeV,
$\Gamma = 260 \pm 40$ MeV.
However, it is at the bottom end of the range of data we analyse,
and we observe some tendency for its mass to drift downwards.
We view the determination of the mass by VES and GAMS as more secure,
but our larger width appears reliable.
If this resonance is dropped from the fit, $\chi ^2$ increases by a definitive
amount, namely 2290.
There are large effects on the fit at many of the low momenta;
effects at two momenta are  illustrated in Figs. 9(j) and (k).

The next $2^+$ state optimises at $2020 \pm 30$ MeV with $\Gamma = 275 \pm
35$ MeV. In earlier fits to $\eta \pi \pi$ data [12] there is the possibility
of cross-talk between the two nearby resonances at 1910 and 2020 MeV, both
lying at the bottom end of
the available mass range in the neutral data.
So the confirmation of $f_2(2020)$ is important.
We find a very large change in $\chi ^2$, namely 2980, when $f_2(2020)$
is removed from the fit and others are re-optimised. Figs. 9(l) and (m)
illustrate the deterioration in the fit to some data.

\begin{figure}
\begin{center}
\vskip -10mm
\epsfig{file=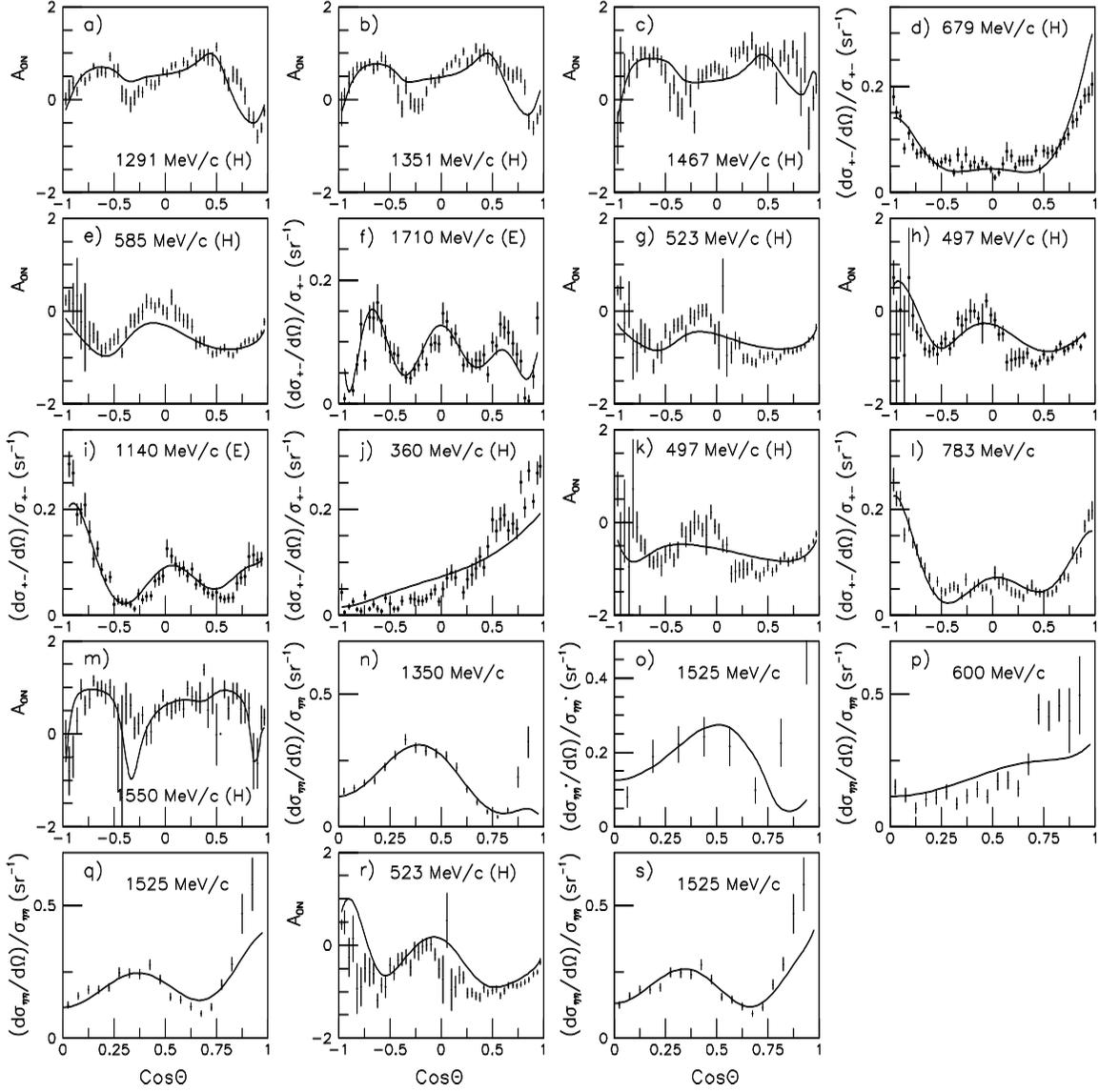,width=18cm}
\vskip -180.46mm
\epsfig{file=NO_RES.PS,width=18cm}
~\
\vskip -6mm
\caption{Illustrating the deterioration in the fits to data when
various resonances are dropped from the fit: (a)--(c) $\rho _5(2295)$,
(d)--(e) $\rho _3(1960)$, (f)--(g) $\rho _3(2210)$,
(h)--(i) $\rho _1(2005)$, (j)--(k) $f_2(1910)$, (l)--(m) $f_2(2020)$,
(n)--(o) $f_2(2230)$ or $f_2(2300)$, (p)--(r) $f_0(2105)$ and (s)
$f_0(2320)$.}
\end{center}
\end{figure}
The present analysis confirms the presence of two further $2^+$ states
at 2230 MeV and 2300 MeV.
When the resonance at 2230 MeV is removed and others are re-optimised,
$\chi ^2$ gets worse by 2290, a very large amount.
The effect of removing it is to produce a fit with a single resonance
at 2280 MeV.
The effect of removing the highest $2^+$ state at 2300 MeV is
2879 in $\chi ^2$.
Again the effect is to move the 2230 MeV resonance close to 2280 MeV and
produce the same bad fit as is illustrated in Figs 9(n) and (o).

The highest of these $f_2$ states, now at $2300 \pm 30$ MeV, was observed in
Refs. [12] and [20] at $2370 \pm 50$ MeV.
This shift in mass arises from a subtle interference with $f_2(2230)$, revealed
by the $\pi ^- \pi ^+$ polarisation data.
For the same reason,
the width of the $f_2(2230)$ is now $245 \pm 45$ MeV, rather than the
$170 \pm 50$ MeV reported earlier.

All four states appear to be dominantly $q\bar q$. We shall discuss below
the flavour mixing angles.

\subsection {$1^- $ states}
The fit is poor without two $1^-$ resonances. The phase advance for
$^3D_1$ on Fig. 6 is evidence for the presence of two states.
However, four are to be expected in
this mass range, originating from $\bar pp$ $^3S_1$ and $^3D_1$.
We cannot identify more than two with confidence.

Figs. 9(h) and (i)  illustrate the effect of dropping the $1^-$ state at
2000 MeV and re-optimising the rest.
There are significant visible
discrepancies in fits to both diffential cross sections and polarisation
at low momenta.
The change in $\chi ^2$ is 1341.
It couples strongly to $\bar pp~^3D_1$ and this makes it distinctive
in polarisation data in the low momentum range where $^3S_1$ dominates.
The $1^-$ state at 2165 MeV has a less significant effect, improving
$\chi ^2 $ by only 450. It couples dominantly to $\bar pp~^3S_1$.
When it is removed from the fit, changes are just discernible by eye.
There are, however, reports of a $1^-$ resonance at 2150 MeV from GAMS data on
the $\omega \pi$ channel [22]. This mass corresponds
closely to what we observe. The width given by GAMS is large,
$320 \pm 70$ MeV. The width we observe is smaller, but has a considerable
error, $160 ^{+140}_{-70}$ MeV, so there is no discrepancy.

We find it is also essential to include a strong $1^-$
contribution peaking at or below the $\bar pp$ threshold. It fits well
as a contribution from $\rho (1700)$. It is possible that some of this effect
originates from the $^3S_1$ threshold in $\bar pp$, but a fit using only
this threshold is somewhat poorer.

We have searched for further $1^-$ resonances at higher mass.
There is some improvement in $\chi ^2$, with an additional state around
$2270$ MeV. However, the analysis will not
support the presence of three $1^-$ states.

\subsection {$0^+$ states}
One of the very striking features of the $\eta \eta$ data is a strong
peak in the integrated cross section of Fig. 5(c) at 2100 MeV. It is not
fitted by $f_4(2020)$ and $f_2(2020)$, whatever the centrifgual barrier radius.
These states with non-zero spin are separated by their characteristic angular
dependence, despite possible interferences between them.
The data demand very strongly the presence of an $f_0$ with mass $2105 \pm 15$
MeV.
When it is dropped from the fit, $\chi ^2$ gets worse by a very large amount,
namely 4030.
The description of the integrated $\eta \eta$ cross section is then much
worse, and there are also discrepancies
with $\eta \eta$ differential cross sections, illustrated with  examples
in Figs. 9(p)--(r). This resonance has the best determined mass and
width of all the
resonances observed in the present analysis, except for $f_4(2020)$.

The mass agrees well with
that observed by the E760 group [27] as a strong peak in $\eta \eta$ in
the reaction $\bar pp \to \eta \eta \pi ^0$ at two beam momenta of 3.1 and
3.5 GeV/c.
The width we fit here, namely $200 \pm 25$ MeV, is close to
that fitted by E760: $203 \pm 10$ MeV.
These parameters  also agree well with the mass fitted to a $0^+$ peak
in the $4\pi$ channel in $J/\Psi \to \gamma (4\pi)$ [28].
This resonance has also been identified as having $J = 0$ in an analysis of
Crystal Barrel data on $\bar pp \to \eta \eta \pi ^0$ [29], where it
appears as a strong peak in $\eta \eta$; there, the determination of mass and
width were not so precise, since the resonance appears at the
top end of the available phase space.
It is presently incorrectly listed by the Particle Data Group
under $f_2(2150)$.

We find a strong requirement for a further $0^+$ resonance at
2320 MeV. If it is omitted, the fit gets visible worse, as is illustrated in
Fig. 9(s); $\chi ^2$ increases by 1257, a highly significant amount.

There has been a report of a further $0^+$ resonance at 2020 MeV with
a width of 400 MeV [30].
Adding this resonance to the fit, $\chi ^2$ improves by 370, which we
regard as significant.
However, because it lies towards the bottom end of the available
data for $\pi ^0\pi ^0$ and $\eta \eta$ and because of interference
with $f_0(2105)$, its mass and width have sizeable errors,
$M = 2005 \pm 30$ MeV, $\Gamma = 305 \pm 50$ MeV.

\begin{figure}
\begin{center}
\vskip -10mm
\epsfig{file=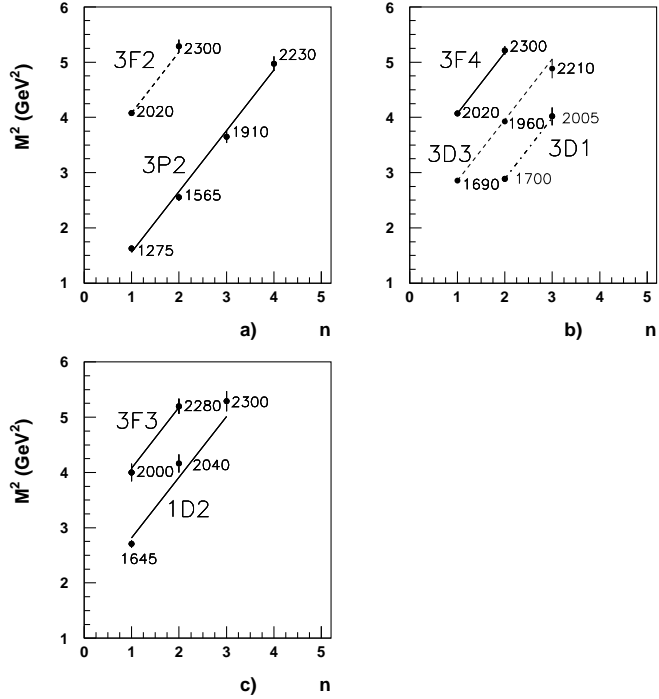,width=10.5cm}
~\
\vskip -6mm
\caption{(a) Suggested trajectories for $^3P_2$ states (full line)
and $^3F_2$ (dashed) v. radial quantum number $n$;
(b) likewise for $4^+$ states (full line),
$3^-$ (dashed) and $1^-$ (dotted);
(c) trajectories for $3^+$ and $2^-$ states.
Numerical values give masses in MeV.}
\end{center}
\end{figure}
\section {Systematics of resonances masses}
On Fig. 10(a), we plot mass squared of $2^+$ resonances against their
recurrence number.
It is possible to construct an almost straight line from
$f_2(1270)$, $f_2(1565)$, $f_2(1910)$ and $f_2(2230)$.This is
illustrated by the full line on Fig. 10(a).
We make use of one result as yet unpublished [31].
The mass of the $f_2(1565)$ has been obtained from an analysis of
Crystal Barrel data on $\bar pp \to \omega \omega \pi ^0$ at rest.
From a fit with a Flatt\' e form, the K-matrix mass is determined to
be $1598 \pm 11(stat) \pm 9(syst)$  MeV.

The $f_2(2020)$ and $f_2(2300)$
are readily placed on a parallel trajectory, shown by the dashed line of
Fig. 10(a). The proximity of the $f_2(2020)$ to $f_4(2020)$ then suggests
strongly that it is the $q\bar q$ $^3F_2$ $n = 1$ state and the
$f_2(2300)$ is its radial excitation. The states on the full line of
Fig. 10(a) are naturally interpreted as $q\bar q$ $^3P_2$ states.
We interpret the shift in mass between the two lines
as originating from the centrifugal barrier in the $q\bar q$
system. This provides an effective repulsion at small radii which
shifts the resonance mass up more for $L = 3$ ($^3F_2)$  than $L = 1$
$^3P_2$.

\begin{figure}
\begin{center}
\vskip -10mm
\epsfig{file=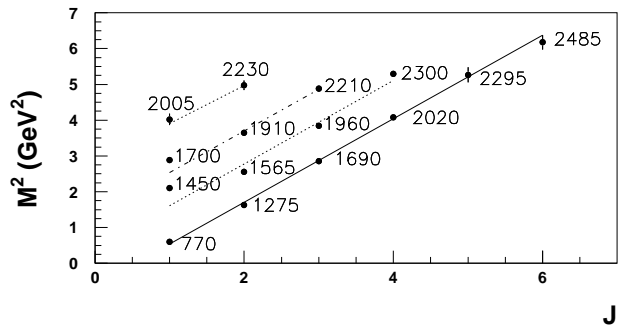,width=10cm}
~\
\vskip -6mm
\caption{Regge trajectories for mesons with spins 1 to 6. Numerical values
give masses in MeV.}
\end{center}
\end{figure}
The slopes of the lines on Fig. 10(a) are 1.10 GeV$^2$, with an
error of $\pm 0.03$ GeV$^2$.
We find that it is possible to construct similar trajectories of
identical slope for other quantum numbers.
Fig. 10(b) shows corresponding trajectories for $4^+$, $3^-$ and $1^-$ states.
There is a well defined shift in mass between $\rho _3(1960)$ and
$f_4(2020)$, which may again be attributed to the effect of the
centrifugal barrier, namely 50--60 MeV per unit of $L$.
The dotted line on Fig. 10(c) shows a possible trajectory
made up of $\rho (1700)$ and $\rho (2005)$.

For $J^P = 2^-$, there is some indication of a discrepancy with a straight
trajectory. However, there is evidence for an extra state $\eta _2(1875)$
[32]. If all these $2^-$ states are confirmed, this is likely to be an intruder
state.
It has a mass roughly that expected for a $2^-$ hybrid [33].

From the analysis of $\eta \pi ^0 \pi ^0$ data, we can identify
corresponding trajectories for $I = 0$ $q\bar q$ $3^+$ and $2^-$
states.
They are shown on Fig. 10(c).
For $3^+$, they are compatible
with a trajectory of the same slope as other $J^P$ on Fig. 10, and with
$^3F_3$ states approximately degenerate with $^3F_4$ and $^3F_2$.
The assignment of $0^+$ resonances is presently controversial, so
we do not show a trajectory for $0^+$.

Fig. 11 shows Regge trajectories. The leading
trajectory made up of $\rho (770)$, $f_2(1270)$,
$\rho _3(1690)$, $f_4(2020)$, $\rho _5(2295)$ and $f_6(2485)$ is well
known.
These resonances lie close to a straight line versus $s$ of slope 1.18 GeV$^2$
per unit of spin.
One can just discern a possible slight difference ($\sim 0.1$ GeV$^2$)
between the
trio $1^-$, $3^-$ and $5^-$ and the even spin states.
The slope is significantly greater than that of the lines on Fig. 10.
Proof of this is the mass difference between $f_4(2020)$ and its
grand-daughter $f_2(1910)$, likewise between $\rho _5(2295)$ and
$\rho _3(2210)$ and the very clearly determined difference
between $f_4(2020)$ and $\rho _3(1960)$.
The centrifugal barrier
in the $\bar qq$ system provides an explanation why high $J$ states move
up in mass.
The intercept at zero mass with the vertical axis is at -0.67, slightly
lower than the accepted value of -0.55 deduced from $\pi N$ scattering data
[34];
however, the systematic shift between even and odd spins on this
trajectory allows the possibility of moving the intercept up by 0.1.

Fig. 11 also shows further parallel Regge trajectories.
It is interesting that $\rho (1700)$ and $\rho (1450)$ depart
significantly from straight-line trajectories through states of
higher spin.
Donnachie and Clegg [35] have argued that decay modes of $\rho (1450)$
require it to be a hybrid or mixed with a hybrid.

\section {Flavour mixing}
We have explained in section 2 that there is a blurred distinction
between (i) allowing phases $\phi _i$ of decay channels to vary from 0 or
(ii) allowing flavour mixing through the angles $\Phi$.
Nonetheless, we feel it is worth summarising the fitted values of
flavour mixing angles $\Phi$ for $2^+$ and $4^+$ states.
These are shown in Table 3.
Errors on individual flavour mixing angles are typically $\pm 5^{\circ}$.
However differences in $\Phi$ between different resonances are strongly
correlated. These differences are what allow an acceptable fit to the
data via small deviations from SU(3).
Nonetheless some flavour mixing is inevitable.
If they are all set to zero, there is an unacceptable increase of
2646 in $\chi ^2$.

Table 3 also includes fitted values of
corresponding ratios $r$ of amplitudes for spins 1, 3,4 and 5.
Errors are typically $\pm 15\%$.
One discerns a trend from positive values for low masses to negative
values at high masses. As the mass rises, the momentum in the
$\bar pp$ channel increases and the first zero of $j_L(kr)$
describing the $\bar pp$ channel moves to smaller radius. As it moves
inside the radius where the resonance wave function peaks, one expects
a change of sign of the ratios of amplitudes $r$ between $L = J \pm 1$,
in qualitative agreement with the observations.
However, because high mass resonances have several nodes in the
wave function of the resonance, it is difficult to be quantitative
for $J^P = 2^+$ and $1^-$ states at present.

\begin{table} [htp]
\begin{center}
\begin{tabular}{ccc}
Resonance & $\Phi (^{\circ})$ & $r_J$\\\hline
$f_2(1910)$ & 1.1 & 2.89\\
$f_2(2020)$ & 7.9 & 1.62 \\
$f_2(2230)$ &  7.5 & -0.38\\
$f_2(2300)$ & -14.8 & -0.92\\
$f_2(2620)$ &  (0) & -0.83\\
$f_4(2020)$ & -26.1  & 0.04 \\
$f_4(2300)$ &  22.9  &  0.35\\\hline
$\rho _5(2295)$ & & (0) \\
$\rho _5(2500)$ & & 0.24 \\
$\rho _3(1960)$ & & 0.04 \\
$\rho _3(2210)$ & & 0.35 \\
$\rho _3(2300)$ & & 2.0 \\
$\rho _1(2005)$ & & 5.0 \\
$\rho _1(2165)$ & &-0.22 \\\hline

\end{tabular}
\caption {Values of flavour mixing angles $\Phi$ and ratios of amplitudes
$r_J = |f_{L = J + 1}|/|f_{L = J-1}|$ as $s \to \infty$.}
\end{center}
\end{table}

A possibility is that the flavour mixing arises purely from the overlap
with neighbouring $s\bar s$ states.
The $s\bar s$ partner of $f_2(1565)$ could be the $2^+$ state reported
by the LASS group at 1950 MeV [36] decaying to $K^*\bar K^*$, in which
case mixing with $f_2(1910)$ would be natural.
On the other hand, the LASS group may simply be observing the
decays of $f_2(1910)$.
The $s\bar s$ partner of $f_2(1910)$ itself is expected
around 2150--2250 MeV.
It is possible to identify the states observed in
$\phi \phi$ as the $s\bar s$ partners of $f_2(1910)$ and $f_2(2020)$.
The $2^+$ state decaying to the $\phi \phi$ S-wave, reported by
Etkin et al. [37], also peaks at about 2150 MeV; however, their K-matrix
analysis assigns it a mass of 2020 MeV, because of a strong threshold effect.
The JETSET group have likewise reported a signal in $\bar pp \to \phi \phi$
peaking at $\sim 2180$ MeV, possibly the same object [38].
Prokoshin has reported a peak in $\eta \eta$ at $2175 \pm 20$ MeV [39],
again possibly the same resonance.
Recent data of the Omega group on central production of $K^+K^-$ reveals
a peak at 2150 MeV [40].

Etkin et al. have also reported a peak in the $\phi \phi$ D-wave at
2300--2340 MeV.
This makes a natural candidate for the $^3F_2$ $s\bar s$ partner of
$f_2(2020)$.

Production of the $\phi \phi$ resonances from initial $\bar pp$ and $\pi
\pi$ states requires explanation. It could arise if resonances are
strongly mixed between $q\bar q$ and $s\bar s$. However, the
small flavour mixing angles we observe do not point that way.

A second clear possibility is that the mixing is with
the $2^+$ glueball predicted in this mass range, or slightly above, by
Lattice QCD calculations.
There is evidence from the Crystal Barrel data on $\bar pp \to \eta \eta \pi
^0$ for a broad $2^+$ state in $\eta \eta$ with a mass of $1980 \pm 50$ MeV
and a width $\Gamma = 500 \pm 100$ MeV [29].
There is similar evidence from two other sources.
Firstly, a broad $2^+$ contribution to the $4\pi$ channel is observed in
central production data at small $p_T$ [30].
Its mass is $1920 \pm 20$ MeV with a width of $450 \pm 60$ MeV.
Secondly, there are data from BES on $J/\Psi \to \gamma {4\pi}$ [41] and
$J/\Psi \to \gamma K^*\bar K^*$ [42], which
both require a broad $2^+$ signal peaking at about 2000 MeV.
The presence of this wide $2^+$ signal in $J/\Psi$ radiative decays and
central production data is
suggestive of mixing between a glueball and neighbouring $q\bar q$ states
to produce a broad state.

We have tried inserting a broad $f_2(1980)$ with $\Gamma = 500$ MeV
into our analysis, in addition to the four $2^+$ states discussed
above. There is a modest improvement in $\chi ^2 $ of $\sim 100$,
but little visible change to the quality of the fits.
This is not sufficient to confirm the possible presence of a
broad $2^+$ background in the present data, because of the complexity of
possible interferences with $f_2(1910)$ to $f_2(2300)$.

\subsection {$f_0(2105)$}
One very striking result of the present analysis is that $f_0(2105)$ is far
from being a normal $q\bar q$ state. Its flavour mixing angle $\Phi$ is
$+(59-64)^{\circ }$ if $f_0(2005)$ is excluded from the fit, or
$+(68-71.6)^{\circ}$ with it included.
That is, it behaves predominantly as an $s\bar s$ state.
Allowing for interferences with $f_0(2005)$, it makes up
$(4.6 \pm 1.5)\%$ of the $\pi ^0\pi ^0 $ $J^P = 0^+$ intensity and
$(38 \pm 5)\%$ of that for $\eta \eta$.
The branching ratio to $\eta \eta '$ is not well determined, because of
the low statistics in this channel.
Our best estimate of amplitude ratios is as follows:
\begin {equation}
\pi ^0 \pi ^0 : \eta \eta : \eta \eta ' = 0.71 \pm 0.17 : 1 : -0.85 \pm 0.45.
\end {equation}
For an unmixed $q\bar q$ state, the ratio expected between $\pi ^0 \pi ^0$ and
$\eta \eta$ is $0.8^{-4}$, i.e. 2.44.
The strong production of an $s\bar s$ state in $\bar pp$ annihilation is
clearly anomalous. How can this be explained?

It could be a second glueball.
A glueball would have a mixing angle of +37$^{\circ }$.
The flavour mixing angle we observe then required some mixing with
a nearby $s\bar s$ state; for ideal mixing, $s\bar s$ states are 
members of SU(3) singlets, so such mixing is plausible.
The latest Lattice QCD calculations [43] predict a second $0^+$ glueball
with a mass ratio to the first of $1.54 \pm 0.11$.
If we assign $f_0(1500)$ and $f_0(2105)$ as the two glueballs,
the mass ratio is 1.40.

The alternative scenario, developed by Anisovich and Sarantsev [44],
is that the glueball mixes strongly with neighbouring $q\bar q$ and
$s\bar s$ states to generate a broad state which contains $\sim 50\%$ of
the glueball in its wave function. Such a broad state may well overlap
the mass range of the $f_0(2105)$. The $f_0(1500)$, $f_0(1770)$ and
$f_0(2105)$ are all seen strongly as peaks in $J/\Psi$ radiative decay
to $4\pi$ [28]. This suggests that $f_0(2105)$ may be a mixed state
made by mixing of the glueball with a nearby $s\bar s$ state.
It is important to look for it in decays to $K\bar K$.

\section {Conclusions}
A combined analysis of the $\pi ^- \pi ^+$ data of Eisenhandler et al.
and Hasan et al. with Crystal Barrel data for
$\pi ^0 \pi ^0$, $\eta \eta$ and $\eta \eta '$
leads to a secure partial wave analysis and the identification
of most of the expected $q\bar q$ resonances in this mass range, except
for two $1^-$ states.
The polarisation data play a vital role in separating states with
$L = J \pm 1$.
The resulting fit is rather close to that obtained previously by
Hasan and Bugg
[9].
However, it is now possible to identify clearly four $2^+$ resonances.
These have small flavour mixing and are probably to be identified with
the $q\bar q$ $^3P_2$ and $^3F_2$ states expected in this mass range.

The systematics of resonances on Fig. 10 establishes a valuable guide
to masses to be expected to $q\bar q$ states.
This should help in identifying intruder states of exotic character.

The $f_0(2105)$ make a large contribution to the $\eta \eta$ and
$\eta \eta '$ data. It is not to be identified with a simple
$q\bar q$ state, because of its +(59--71.6)$^{\circ }$ flavour mixing angle.
This requires it to have a large $s\bar s$ and/or exotic component.

\section{Acknowledgement}
We thank the Crystal Barrel Collaboration for
allowing use of the data.
We acknowledge financial support from the British Particle Physics and
Astronomy Research Council (PPARC).
We wish to thank Prof. V. V. Anisovich for helpful discussions.
The St. Petersburg group wishes to acknowledge financial support from PPARC and
INTAS grant RFBR 95-0267.
We wish to thank Dr. D. Peaslee for helpful discussions on the data
of Dulude et al, and also concerning Regge trajectories.

\end {document}